\documentclass[%
prl,
twocolumn,
superscriptaddress,
amsmath,amssymb,
 aps,
]{revtex4-2}
\usepackage{xcolor}
\usepackage{url}
\usepackage{graphicx}% Include figure files
\usepackage{dcolumn}% Align table columns on decimal point
\usepackage{bm}% bold math
\usepackage[separate-uncertainty = true]{siunitx}
\usepackage{booktabs, siunitx, graphicx}
\begin{document}

\title{\textbf{Spreading and absorption of silicone oil droplets on silicone elastomer films} 
}% 

\author{Lauren Dutcher}
\affiliation{Department of Physics and Astronomy, McMaster University, Hamilton, Ontario, Canada.}
\author{Benjamin Baylis}
\affiliation{Department of Physics, University of Guelph, Guelph, Ontario, Canada.}
\author{John R. Dutcher}
\affiliation{Department of Physics, University of Guelph, Guelph, Ontario, Canada.}
\author{\'Elie Rapha\"el}%
\affiliation{UMR CNRS Gulliver 7083, ESPCI Paris, PSL Research University, Paris, 75005, France.}
\author{Kari Dalnoki-Veress}
 \email{dalnoki@mcmaster.ca}
\affiliation{Department of Physics and Astronomy, McMaster University, Hamilton, Ontario, Canada.}
\affiliation{UMR CNRS Gulliver 7083, ESPCI Paris, PSL Research University, Paris, 75005, France.}

\date{\today}% It is always \today, today,

\begin{abstract}
When a liquid droplet completely wets a hard substrate, its spreading dynamics follow Tanner’s law, with the droplet radius growing as the one-tenth power of time. Here, we investigate how these dynamics change when silicone oil droplets spread on soft silicone elastomer and gel films supported by a rigid silicon substrate. While the droplets fully wet the elastomer surface, they also simultaneously swell the elastomer film. By varying the film thickness, we observe deviations from the classical power-law scaling, which we interpret in terms of changes to the effective stiffness and the absorption potential of the system. We describe the spreading behavior using a phenomenological model that accounts for both absorption and mechanical contributions.
\end{abstract}

                              
\maketitle

\section*{Introduction}

Liquid droplets interact with surfaces in complex ways, influenced by surface chemistry and bulk substrate properties. Although simple cases such as liquids on solid substrates, immiscible liquid substrates, and even permeable substrates have been well characterized, more complex systems involving simultaneous deformation and absorption remain largely unexplored~\cite{velde2023spreading}.

Substrate wettability plays an important role in industrial applications like ink jet printing~\cite{printing1, printing2}, in biological processes involving plant leaves such as rainwater collection~\cite{plants} and pesticide use~\cite{pesticide}, and is utilized in microfluidic devices~\cite{microfluidics1,microfluidics2}. For a partially wetting liquid, an equilibrium contact angle $\theta_\text{Y}$ forms between the materials.  For a rigid substrate, Young's contact angle is given by $\cos\theta_\text{Y} = (\gamma_\text{SV} - \gamma_\text{SL}) / \gamma$, where $\gamma_\text{SV}$ is the solid-air surface tension, $\gamma_\text{SL}$ is the solid-liquid surface tension and $\gamma$ is the liquid-air surface tension~\cite{contact_line, contact_angle, hydrodyn_cont_ang, Duffy}. When a liquid completely wets a substrate, $\theta_\text{Y} =0$, and the droplet eventually forms a thin film across the substrate's surface at equilibrium~\cite{degennes_text}.

%Substrate wettability plays an important role in industrial applications like ink jet printing~\cite{printing1, printing2}, in biological processes involving plant leaves such as rainwater collection~\cite{plants} and pesticide use~\cite{pesticide} and is utilized in microfluidic devices~\cite{microfluidics1,microfluidics2}. The spreading parameter: $S = \gamma_\text{SV} - \gamma_\text{SL} - \gamma$ characterizes the wettability of a surface with a given liquid, where $\gamma_\text{SV}$ is the solid-air surface tension, $\gamma_\text{SL}$ is the solid-liquid surface tension and $\gamma$ is the liquid-air surface tension. For a partially wetting liquid with a negative spreading parameter, a contact angle $\theta_\text{Y}$ forms between the materials.  For a rigid substrate, this contact angle, given by $\cos\theta_\text{Y} = (\gamma_\text{SV} - \gamma_\text{SL}) / \gamma$, is well understood~\cite{contact_line, contact_angle, hydrodyn_cont_ang, Duffy}. A positive spreading parameter signifies that the liquid completely wets the substrate, $\theta_\text{Y} =0$, forming a thin film across the substrate's surface~\cite{degennes_text}.

The spreading dynamics is generally well captured at late stages by a power law where the contact radius varies with time as $R\propto t^m$~\cite{deGennes, Verneuil, Roques}. The case of a liquid droplet that completely wets a rigid substrate has been well characterized~\cite{deGennes, Starov, Verneuil, Bird, Snoeijer, Complex_wet} and is given by Tanner's law when viscous dissipation dominates (i.e. we can ignore gravitational and inertial effects). Tanner's law describes the radial growth: $R \sim t^{1/10}$~\cite{Tanner, degennes_text}, and holds for late stage spreading in Newtonian fluids where gravitational and inertial effects can be neglected. The 1/10th power law is material independent as a precursor film extends ahead of the droplet's wetting front such that the bulk droplet spreads across a nanometrically thin film of itself~\cite{precursor}. The slow power law dynamics can be attributed to the fact that the precursor film that precedes the spreading of the bulk droplet, eliminates the reduction in the free energy associated with coating the solid-air interface; rather, it is only the Laplace pressure associated with the curvature of the droplet that drives the wetting front forward. The prefactor preceding the 1/10th scaling captures properties  such as the viscosity, $\eta$, the surface tension of the liquid, $\gamma$, and droplet volume,~$\Omega$. 

In addition to droplets spreading on solid substrates, spreading of liquid droplets on liquid substrates have also been well studied~\cite{cormier, liquid_sub, liquid_sub2, liquid_sub3}. If a liquid droplet is placed on a soft solid instead, deformations in the substrate modify the wetting dynamics. This deformation, known as a capillary ridge, arises from the liquid’s surface tension, $\gamma$, acting at the contact line in opposition to the substrate’s elastic modulus, $E$~\cite{Andreotti, Chen, Hauer,cai2021fluid, Zhao}. The length scale that characterises this deformation is the elastocapillary length $l_{ec} = \gamma/E$~\cite{energy, review, review2}. The relative energy dissipation between the liquid and the deformable solid determine how the spreading dynamics diverge from Tanner's law; and, the energy dissipated in the capillary ridge causes `visco-elastic' braking \cite{first, adhesion, spreading, indepen_vis, surf, Long,cai2021fluid}. In contrast, Tamin and Bostwick have theoretically shown that a viscous drop placed on a purely elastic substrate can have faster spreading dynamics on softer films due to contact angle changes~\cite{thin_drop}, while viscoelasticity decreases the spreading rate.

% soft solid examples: (like biological membranes, etc etc *cite*)

Introducing a permeable material further alters the wetting dynamics. Porous substrates are typically characterized by their pore geometry as well as the capillary pressure, which drives absorption and  depends on surface tension, contact angle and pore size~\cite{pores}. When a liquid droplet wets a porous substrate, two competing processes impact the dynamics: the spreading of the liquid atop the substrate increasing the contact area and the absorption of the liquid into the substrate causes a volume loss in the droplet, thereby decreasing the contact area~\cite{velde2023spreading, imbibition1, imbibition2, imbibition3, imbibition4, porous, Seveno}. Initially, the wetting process dominates and the droplet radius grows. As the contact area increases, so does the absorption potential therefore spreading slows down until a maximum contact area is reached. Finally, given enough absorption into the substrate the droplet radius recedes~\cite{imbibition1, imbibition2, imbibition3, imbibition4, porous}. Spreading dynamics on porous substrates have been studied both experimentally and numerically~\cite{Clarke, Frank, Johnson, Chen_Nie, Chebbi, Concalves, Moreton}. 

Van de Velde and co-workers investigated the coupled spreading and absorption of droplets on semi-infinite gels and as-prepared elastomeric surfaces~\cite{velde2023spreading}. Our study builds on their work and on that of Hourlier-Fargette and co-workers~\cite{hourlier2017}, who showed that as-prepared elastomers exhibit complex wetting behaviour due to uncrosslinked chains remaining after fabrication. These chains swell the network, imparting gel-like characteristics rather than those of a true elastomer~\cite{cai2021fluid,hourlier2017,Hamza_droplet,Hamza_pdms}, and can account for up to 30\% of the total volume in as-prepared samples~\cite{Hamza_pdms}.

Here, we investigate droplets on true elastomeric substrates that both deform and swell in the presence of the droplet, using silicone oil droplets on crosslinked silicone films. As in the work of Van de Velde and co-workers, spreading and absorption occur simultaneously. To disentangle the effects of absorption and substrate mechanics, we vary the film thickness while keeping a rigid support beneath. This geometry has two principal consequences: (1) the \emph{effective} stiffness changes with thickness, just as a thin mattress on a hard surface feels stiffer than a thicker one of the same material; and, (2) the absorption potential is tuned, with thin films taking up less liquid than thick films. In the limit of  small film thickness, both deformation and absorption make negligible contributions to the spreading dynamics.

We track the spreading with optical microscopy and analyse it using an empirically motivated model that quantifies the coupled influence of mechanical deformation and absorption on the dynamics.

\section*{Experimental Methods}

 We use PDMS (Gelest, DMS-V35 and DMS-V51, with approximate viscosities of 5000~cSt and 100000~cSt) with a PDMS based co-polymer (Gelest, HMS-064, approximate viscosity 6,000-9,000 cSt) in a weight ratio of 20:1 dissolved in toluene to a total polymer concentration of 25\% (weight/weight). This 25\% stock solution is further diluted down to solutions with a total polymer concentration as low as  0.5\%. To 1~g of the diluted PDMS/co-polymer solution, 10 µL of 400x diluted catalyst Platinum(0)-1,3-divinyl-1,1,3,3-tetramethyldisiloxane complex (Sigma Aldrich) is added to promote crosslinking prior to spincoating the elastomer films. This concentration of catalyst ensured that there is always an excess for all solution concentrations used. A few drops of this solution are placed onto 1 cm x 1 cm cleaved Silicon wafers and spincoated at 3500 rpm for 30 s (Speedline Technologies) to produce thin, uniform gel films. The solution concentration is adjusted to obtain desired film thicknesses. After spincoating, the films are placed in an oven at 85\textdegree C for $\sim$ 24 h to cure the films. 

\begin{figure}[]
\centering
\includegraphics[width=220 pt]{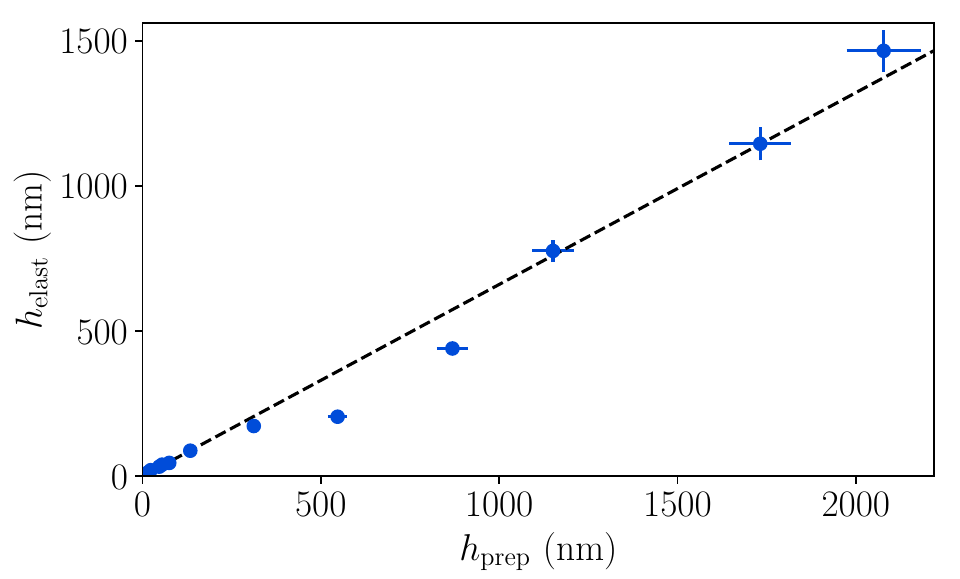}
\caption{
Film thickness measured before and after the washing procedure. The as-prepared film thickness is $h_{\text{prep}}$, while $h_{\text{elast}}$ is the thickness of the elastomer film following washing. The data is fit to a straight line with a best-fit slope of  0.66. Error bars represent typical 5\% uncertainty.}
\label{fig:rinse}    
\end{figure}

To make the elastomer films we complete a series of solvent wash steps which involve placing a droplet of toluene on the film for 60 s, then spinning at 3500 rpm for 15 s to remove the toluene and all uncrosslinked chains. This 60 s wash step is done in triplicate. The films are placed in baths of toluene overnight to further ensure removal of the uncrosslinked chains, and complete the preparation of the elastomer films~\cite{Hamza_pdms}. 
Again, we emphasize that rinsing removes a significant fraction of uncrosslinked chains, as shown in Fig.~\ref{fig:rinse}, which plots the post-rinse elastomer thickness versus the as-prepared thickness for crosslinked DMS-V35 films. The relationship between the elastomer and as-prepared film thickness obtained from  Fig.~\ref{fig:rinse} is well described by $h_{\text{elast}}\approx 0.66 h_{\text{prep}}$.

Spreading dynamics were investigated on several types of samples. To examine the role of the elastomer modulus, films were prepared from DMS-V35 and DMS-V51. The higher viscosity DMS-V51 has a higher molecular weight, resulting in longer chain segments between crosslinks and a lower modulus. The modulus of the elastomers was measured by atomic force microscopy (AFM) force spectroscopy on films of thickness $h \sim 1000$~nm, using an indentation depth of $\sim h/10$. Force--distance curves were analyzed with the Hertz model for a pyramidal tip to obtain the modulus $E$, assuming a Poisson's ratio of $1/2$ (appropriate for an incompressible solid such as an elastomer). We obtained $E(\mathrm{V51}) = \SI{790 \pm 70}{\kilo\pascal}$ and $E(\mathrm{V35}) = \SI{1230 \pm 110}{\kilo\pascal}$. Although the difference in modulus is modest, we refer to these as the “high modulus” and “low modulus” films. To investigate the difference between gel films and elastomer films, droplet spreading was also measured for as-prepared gel films and elastomer films of the higher modulus DMS-V35 system. These as-prepared DMS-V35 films are referred to as “gel films”. For the modulus of the gel films we use $E(\text{gel)} \approx \SI{500}{\kilo\pascal}$, as measured for the same system previously~\cite{Hamza_droplet}.  

\begin{figure}[]
\centering
\includegraphics[width=220 pt]{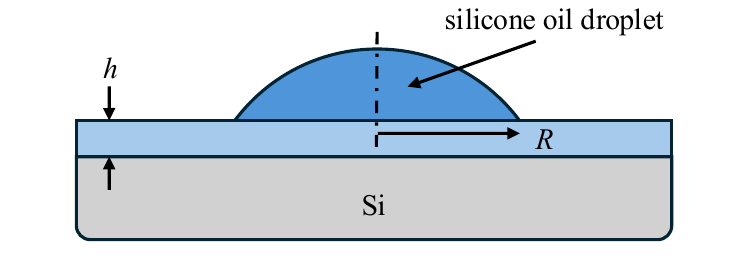}
\caption{
Schematic of the sample geometry. Silicone oil droplet spreads on a film with thickness $h$, supported on a silicon substrate. The film consists of either a gel or an elastomer (see text). The contact radius of the droplet, $R$, is measured as a function of time.}
\label{fig:schematic}    
\end{figure}

Lastly, droplet spreading was measured on a Si substrate to obtain the limiting case $h \rightarrow 0$, corresponding to the ideal case of Tanner’s law wetting. In summary, a total of 31 droplet spreading experiments were carried out on four sample types: (1) high modulus elastomer films ($42$~nm $< h < 1575$~nm) from DMS-V35, (2) gel films ($23$~nm $< h < 2078$~nm) from as-prepared DMS-V35, (3) low modulus elastomer films ($22$~nm $< h < 2560$~nm) from DMS-V51, and (4) a Si substrate ($h \rightarrow 0$). See Figure~\ref{fig:schematic} for a schematic of the sample geometry used.

We quantify  the spreading dynamics of liquid droplets on these substrates by placing a droplet of silicone oil (Gelest, DMS-T35, 5000 cSt, $\gamma=\SI{21.3}{\milli\newton/\meter}$) on top of them. To observe these droplets under an optical microscope, they must be small and deposited to form a near-circular contact area. To achieve this geometry, we use a thin glass rod (diameter $\sim 20 \mu$m), dipped into a bath of the silicone oil and slowly pulled out vertically. A Landau-Levich film~\cite{degennes_text, Maleki} coats the surface of the glass rod, which within seconds, breaks up into droplets via the Plateau-Rayleigh instability~\cite{degennes_text}. The glass rod is then lowered onto the PDMS film and the droplet at the tip of the glass rod is touched to the film and then retracted. This process results in a small droplet that spreads across the surface. 

\begin{figure}[h]
\centering
\includegraphics[width=220 pt]{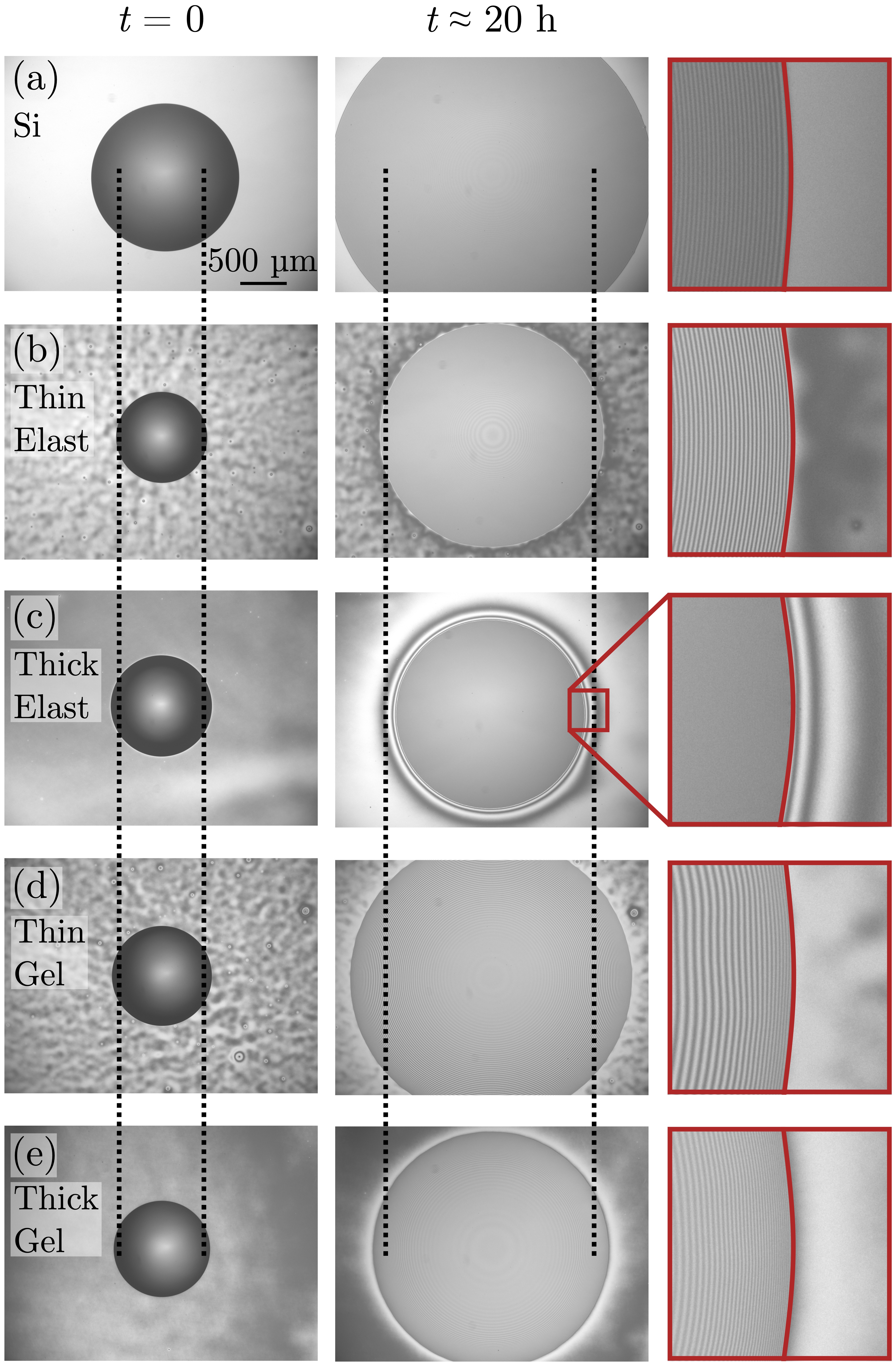}
\caption{Microscopy images of silicone oil droplets wetting different surfaces at \(t=0\) and \(t\approx 20~\mathrm{h}\): (a) bare silicon, (b) a \(232~\mathrm{nm}\) elastomer film supported on silicon, (c) a \(1575~\mathrm{nm}\) elastomer film supported on silicon, (d) a \(312~\mathrm{nm}\) gel film supported on silicon, and (e) a \(2078~\mathrm{nm}\) gel film supported on silicon. Vertical dashed lines are to facilitate comparing droplet radii between samples. The contact line of each droplet is highlighted with a red line in the third column.
}
\label{fig:microscopy}    
\end{figure}

The experiments are completed on a hot stage (Linkam, UK) at 35\textdegree C to ensure consistent temperature, and therefore a uniform liquid viscosity for all experiments. The experiments are viewed from above using a bright-field optical microscope with a 5x objective and a green laser line filter (532 nm, Newport Corp.). The laser-line filter ensures sharp interference fringes which are used to obtain height information to extract droplet volume data. Images were acquired over 20 h at geometrically increasing time intervals, with each interval 1.02 times longer than the preceding one. This corresponds to time points that are approximately evenly spaced on a logarithmic time scale. Figure \ref{fig:microscopy} shows microscopy images for wetting on a Si substrate (a), two elastomer films (thin/thick) supported on Si (b,c), and two gel films (thin/thick) supported on Si (d,e), at their first time frame and their last, 20 h later. Some of the thin films (Fig.~\ref{fig:microscopy}b and d) show a mottled pattern on the substrate away from the droplet. This pattern arises from a long-wavelength thickness non-uniformity introduced during spin coating. Thin-film interference, accentuated by the laser-line filter, makes the pattern visible; the films themselves can be considered uniform: the peak-to-valley height variation is at most \SI{100}{\nano\meter}, with a lateral peak-to-peak spacing of \SI{100}{\micro\meter}.

\section*{Experimental Results}
 
\begin{figure}[]
\centering
\includegraphics[width=235 pt]{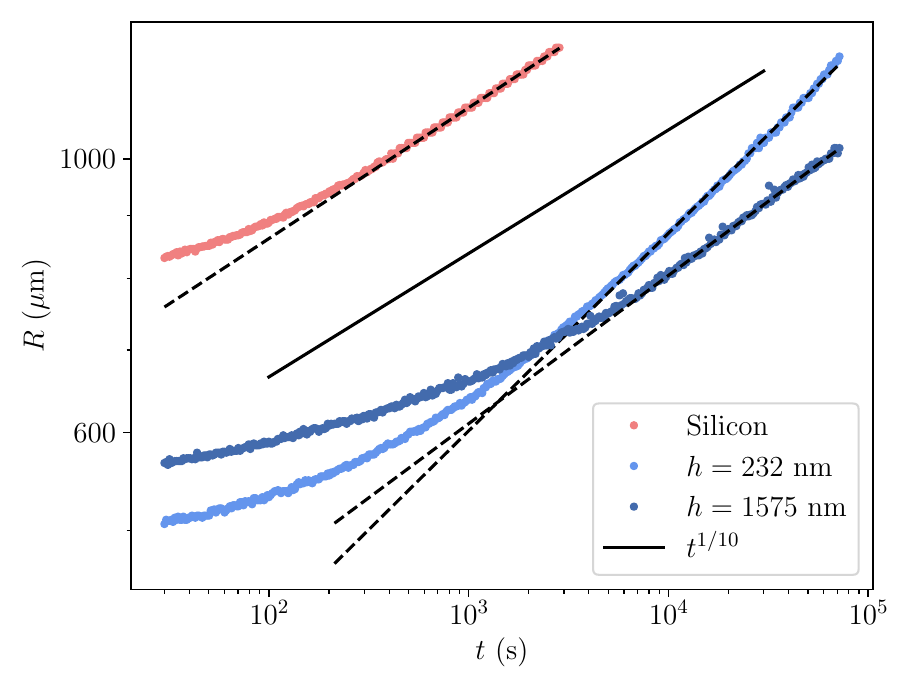}
\caption{Contact radius evolution for droplets of  droplet of silicone oil spreading on two different V35 elastomer film thicknesses, and on a Si substrate (see Fig.~\ref{fig:schematic}). On the Si substrate, the droplet follows Tanner’s law, with the solid black line indicating a $t^{1/10}$ power law. In contrast, droplets on elastomer films exhibit modified dynamics that depend on the film thickness, reflecting the influence of substrate compliance on the spreading dynamics. }
\label{fig:radius}    
\end{figure}

\begin{figure}[]
\centering
\includegraphics[width=235 pt]{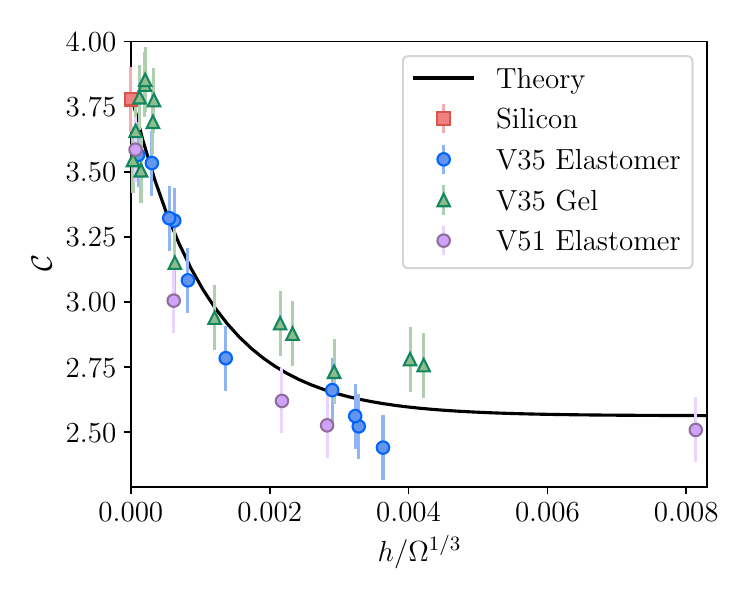}
\caption{Plot of the prefactor, $C$, as a function of the  film thicknesses normalized by the size of the droplet $\Omega^{1/3}$ for elastomer (circles) and gel (triangles) films. The black line shows the  fit of the model described by Equation~(\ref{eqn:prefactor}) to the data. Uncertainties were estimated from the scatter in the data. }
\label{fig:prefactor}    
\end{figure}

\begin{figure}[]
\centering
\includegraphics[width=235 pt]{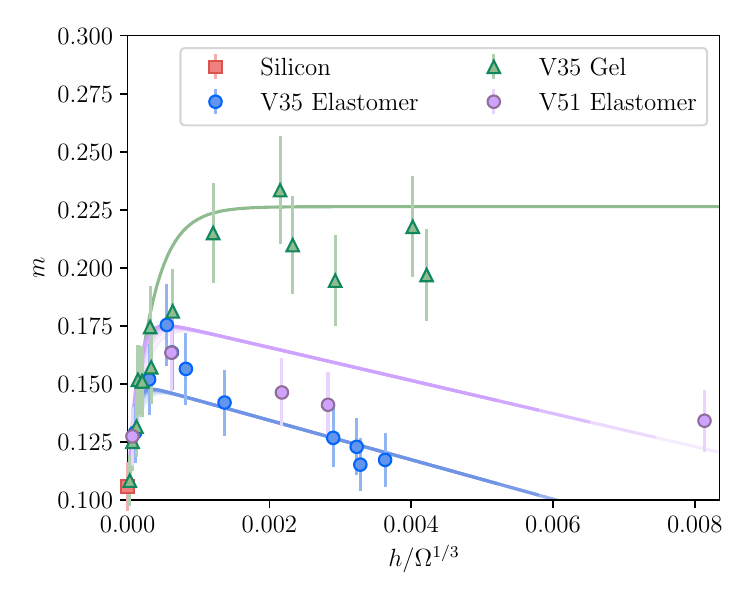}
\caption{Power law exponent, $m$, as a function of the film thicknesses normalized by the droplet size, $\Omega^{1/3}$, for elastomer and gel films. The coloured lines show the model described by Equation (\ref{eqn:m}) for various droplet volumes that correspond to $\Omega^{1/3}=300~\mu$m (lightest) to $\Omega^{1/3}=500~\mu$m (darkest) in increments of $50~\mu$m, which reflects the range of droplet volumes used in the experiments. Uncertainties were estimated from the scatter in the data.}
\label{fig:power}    
\end{figure}

To investigate the dynamics of the spreading event, we measured the silicone oil droplet radius over time. The droplet radius was measured using a Python Hough circle fitting protocol. Figure \ref{fig:radius} shows the typical evolution of droplet radius measured for two different sample thicknesses as well as a Si substrate. We see from Figure \ref{fig:radius} that, as expected, the late stage spreading is well described by a power law~\cite{Verneuil,velde2023spreading, precursor, Chen, liquid_sub, liquid_sub2, liquid_sub3, Bird}. We fit the data to  
\begin{equation} \label{datafit}
    \frac{R}{\Omega^{1/3}} = \mathcal{C} \left( \frac{t}{\tau} \right)^m,
\end{equation}
where $\Omega$ is the droplet volume, $\tau = \Omega^{1/3} \eta / \gamma$ is a characteristic timescale based on the capillary velocity $\gamma / \eta$, and $\mathcal{C}$ is a dimensionless constant.
We determine the volume of of the droplet (assuming a spherical cap) from the droplet contact radius and the dynamic contact angle, $\theta$. The droplet volume is obtained at the  end of the experiments and measured from optical images, using interference fringes at the contact line to determine $\theta$. Typical contact angles at this late stage are $\theta \sim  \SI{0.05}{\radian}$. Eq.~\ref{datafit} has two fitting parameters, the prefactor, $\mathcal{C}$, and the exponent $m$. 
The fit values for $\mathcal{C}$ and $m$ as a function of the film substrate thickness are shown in Figures~\ref{fig:prefactor} and \ref{fig:power} for the four types of substrates. 

% volume equation: $\theta$: $\Omega = \frac{\pi}{4}R^3\theta$

Figures~\ref{fig:prefactor} and \ref{fig:power} show a clear dependence of both the prefactor and the exponent on $h/\Omega^{1/3}$, and on the sample type: elastomer or gel. In the limit $h \rightarrow 0$, the two elastomer cases and the gel converge to the same values of prefactor and exponent, which approach those for droplet spreading on a Si substrate, consistent with Tanner’s law ($m \approx 1/10$)~\cite{Tanner, degennes_text}. 

Comparing the two elastomer thicknesses in Fig.~\ref{fig:radius} reveals that droplet dynamics depend not only on the substrate being an elastomer, but also on the thickness of the elastomer layer, $h$ (see Fig.~\ref{fig:schematic}). The non-monotonic dependence of the power-law exponent (Fig.~\ref{fig:power}) indicates that at least two mechanisms are dominant: (1) \textit{mechanical coupling} — substrate stiffness influences the extent of the capillary ridge and viscous braking, and (2) \textit{absorption} — the elastomer absorbs liquid into the film. Both the effective stiffness and the absorption potential depend on the elastomer film thickness.

A gel film is already swollen by uncrosslinked chains, and while some entropic mixing of droplet chains into the gel, and uncrosslinked gel chains into the droplet, can occur, any additional swelling is much less than that for an elastomer. This is evident in Fig.~\ref{fig:microscopy} in the late stage on the right: in panel (c), for a thick elastomer film, significant swelling is visible from fluctuating concentric interference rings surrounding the silicone oil droplet (the innermost ring), whereas in panel (e), for a gel film of similar thickness, such fringes are absent. This difference in swelling behaviour between elastomer and gel films was consistent across all samples.

Qualitatively, the results can be interpreted as follows. The prefactor, $\mathcal{C}$, decreases monotonically with increasing thickness due to enhanced viscous braking as the substrate becomes more compliant. The similar thickness dependence of $\mathcal{C}$ for high modulus, low modulus, and gel films suggests that, at the contact line, elastomer films are swollen to the point of resembling gels. This implies that local absorption at the contact line is faster than the motion of the contact line itself; indeed, in Fig.~\ref{fig:microscopy}(c), swelling is visible ahead of the contact line. The low modulus data in Fig.~\ref{fig:prefactor} show a slightly larger decrease in $\mathcal{C}$ compared to the high modulus films, but because the prefactor’s sensitivity to modulus is weak, for simplicity we treat its dependence on $h$ as the same for all films.

Turning to the exponent $m$ in Fig.~\ref{fig:power}, again the two dominant effects are: (1) mechanical coupling, which affects spreading on both gels and elastomers, and depends on $h$ and $E$; and (2) absorption, which affects only elastomers, and depends on $h$ since film thickness controls the absorbable volume. As $h$ decreases, the absorption potential vanishes, and all three substrates show similar behaviour for $h \lesssim \SI{100}{\nano\meter}$, where mechanical coupling dominates. Gel films exhibit a monotonic increase in $m$ that saturates above a certain thickness, indicating that beyond this point the deformation does not reach the Si substrate — the droplet no longer senses the underlying Si and the film is effectively semi-infinite. For elastomer films, $m(h)$ is non-monotonic: in addition to mechanical coupling, absorption becomes more significant with thickness, decreasing $m$. Even for the thickest elastomers, the absorbed liquid volume is at most 10\% of the droplet volume (an upper limit assuming complete saturation of the contact area to 30\% of the film’s volume).

\section*{Empirical Model}

Having established that two mechanisms are important to the spreading dynamics, and that the data at long times can be expressed \textit{empirically} as a power-law, we now proceed by expanding the general form of Eq.~\ref{datafit}.
We can write Eq.~\ref{datafit} as
\begin{equation}
\frac{R}{\Omega^{1/3}}=\mathcal{C} \underbrace{\left[ \left(\frac{t}{\tau}\right)^{1/10}\right]}_{\text{Tanner}} \underbrace{\left[ \left(\frac{t}{\tau}\right)^{\mathcal{A}}\right]}_{\text{Absorption}} 
\underbrace{\left[ \left(\frac{t}{\tau}\right)^{\mathcal{M}}\right]}_{\text{Mechanical}},
\label{eqn:fullscaling}
\end{equation}
where Tanner’s law is modified to include an absorption term with exponent $\mathcal{A}$, and a mechanical coupling term with exponent $\mathcal{M}$. Here we have made the assumption that absorption and the mechanical coupling are the two mechanisms that are dominant. In the case of the gel films, the absorption mechanism is not present and $\mathcal{A}=0$.

We first turn to the prefactor, $\mathcal{C}$. From the data in Fig.~\ref{fig:prefactor} we see that in the limit as $h \rightarrow 0$, $\mathcal{C}  \equiv \mathcal{C}_0 \approx 3.8$ and $\mathcal{C}$ is a function that decreases with increasing $h$. We expect this decrease in $\mathcal{C}$ due to energy dissipated in the substrate (viscous braking). We also expect the dependence on $h$ to saturate, since at some thickness the droplet contact line cannot ``sense'' the buried Si substrate. Since $\mathcal{C}$ is dimensionless, $h$ must be non-dimensionalised. The quantity $h/\Omega^{1/3}$ is a natural
choice and hence the prefactor can be written as $\mathcal{C}(h/\Omega^{1/3})$. An exponential asymptotic behaviour is intuitive and we suggest: 
\begin{equation} \label{eqn:prefactor}
    \mathcal{C} = \mathcal{C}_0 + \Delta \mathcal{C}\left(1-e^{-\beta \frac{h}{\Omega^{1/3}}}\right),
\end{equation}
where $\Delta\mathcal{C}$ and $\beta$ are dimensionless fit parameters. The black line in Figure \ref{fig:prefactor} is a best fit of Eq.~(\ref{eqn:prefactor}) to the data with $\Delta\mathcal{C}= -1.24 \pm 0.07$ and $\beta = 900 \pm 150$, which well describes the data.

The mechanical coupling between the droplet and the soft substrate is expected to be a function of $h$, $E$, and $\gamma$, which combines to give us a dimensionless number $hE/\gamma$. Furthermore, eventually the effect of increasing film thickness saturates and a plateau is reached when the droplet does not ``sense'' the underlying Si substrate. This saturation depends on the elastocapillary length ($\gamma/E$) and should appear when $h\gg\gamma/E$. We can assume the mechanical exponent to have the form:
\begin{equation} \label{eqn:mech}
    \mathcal{M}(h, E, \gamma) = \mathcal{M}_\infty (1-e^{-\alpha \frac{hE}{\gamma}}),
\end{equation}
where $\alpha$ is a fit parameter, and $\mathcal{M}_\infty$ is a dimensionless prefactor that must increase with decreasing $E$, since a lower modulus results in a larger plateau value (the softer the film, the bigger the effect). Since the value $\mathcal{M}$ at large $h$ is independent of $h$, $\mathcal{M}_\infty$ cannot have an $h$ dependence. Thus we assume a linear dependence and suggest that $\mathcal{M}_\infty = E_0/E$, where $E_0$ is a fit parameter with the units of a modulus.

The effect of absorption within the system must also depend on the substrate film thickness: in the limit of $h\rightarrow 0$ the effect of absorption must vanish, while in the case of $h \rightarrow \infty$ there is a regime where a droplet can stop spreading and even shrink due to the loss of volume as the liquid absorbs into the film (see the work by \cite{velde2023spreading} as well as~\cite{imbibition1, imbibition2, imbibition3, imbibition4}). Here we are in a regime where even for the thickest films and at the longest times, the loss of volume is at most 10\%. The thickness of the film, into which liquid absorbs, determines the contribution of absorption. Also, for a given thickness, absorption has a larger effect on a small droplet than a large droplet. Hence, the dimensionless quantity $h/\Omega^{1/3}$ sets the scale
for the importance of absorption. Since for all experiments $h/\Omega^{1/3} < 0.01$ we can linearize the dependence of the function $\mathcal{A}$ to:
\begin{equation} \label{eqn:imb}
    \mathcal{A}(h,\Omega) = \lambda \frac{h}{\Omega^{1/3}},
\end{equation}
where $\lambda$ is a fitting parameter. 

\begin{figure}[]
\centering
\includegraphics[width=\columnwidth]{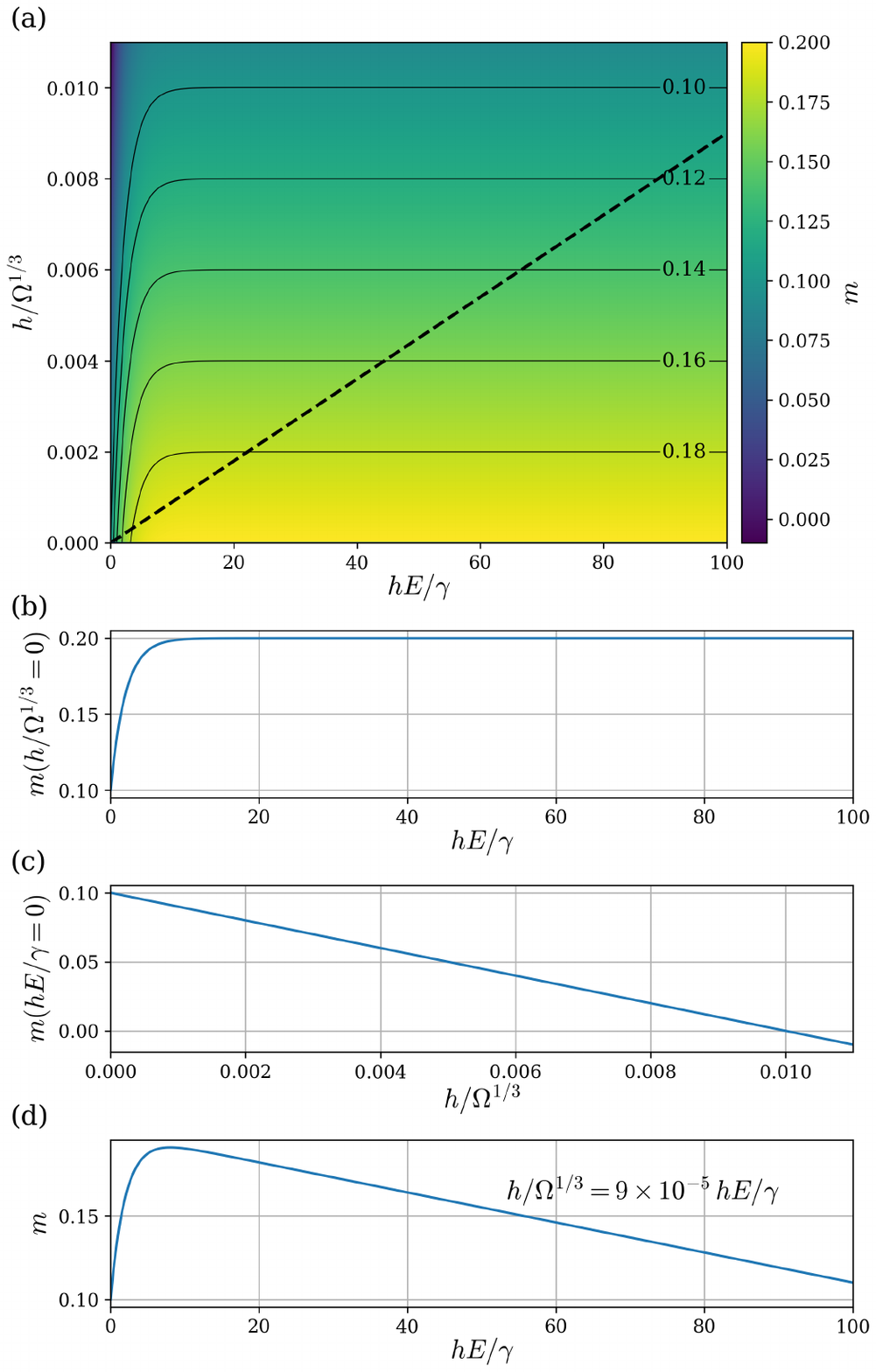}
\caption{Dependence of the exponent $m$ on the two non-dimensionalised film-thicknesses ${hE}/{\gamma}$ (mechanical coupling) and ${h}/{\Omega^{1/3}}$ (absorption). Here the following parameters are chosen to be similar to that of the experiments: $\mathcal{M}_\infty=0.1$, $\alpha=0.5$, and $\lambda=-10$. (a) Contour plot showing $m$ as a function of ${hE}/{\gamma}$ and ${h}/{\Omega^{1/3}}$. (b) $m$ as a function of ${hE}/{\gamma}$ with ${h}/{\Omega^{1/3}}=0$. This is the case of a horizontal slice at the origin and corresponds to no absorption, only mechanical coupling. (c) $m$ as a function of ${h}/{\Omega^{1/3}}$ with ${hE}/{\gamma}=0$. This is the case of a vertical slice at the origin and corresponds to no mechanical coupling, only absorption. (d) $m$ as a function of ${hE}/{\gamma}$, with ${h}/{\Omega^{1/3}}=9 \times 10^{-5} {hE}/{\gamma}$. This plot shows how the exponent varies along the black dashed line in (a), and corresponds to a typical experimental curve for an elastomer where both absorption and mechanical coupling are important with the typical experimental values of $\Omega=300~\mu$m and $\gamma= \SI{21.3}{\milli\newton\per\meter}$ and $E=\SI{790}{\kilo\pascal}$. Note the similarity in this panel to the data in Fig.~\ref{fig:power}. 
 }
\label{fig:m}    
\end{figure}

From Equations~(\ref{eqn:fullscaling}), (\ref{eqn:mech}), and (\ref{eqn:imb}), we have 
\begin{equation}
    m=\frac{1}{10}+\frac{E_0}{E} \left(1-e^{-\alpha \frac{hE}{\gamma}}\right)+ \lambda \frac{h}{\Omega^{1/3}}.
    \label{eqn:m}
\end{equation}
Here $E_0$, $\alpha$, and $\lambda$ are the only free parameters. Having taken into account the material properties,  $E_0$  and  $\alpha$ should be the same for all samples. In contrast, $\lambda=0$ for the gel samples since there is no significant absorption, and since absorption can be affected by the cross-link density of the elastomers, we expect some variation in $\lambda$ for the two types of elastomers. Before we review the fit of the model to the data, we first investigate Eq.~\ref{eqn:m}. Here ${hE}/{\gamma}$ and ${h}/{\Omega^{1/3}}$ are the non-dimensional film thickness for the mechanical coupling term and the absorption term. In Fig.~\ref{fig:m} we plot $m$ as a function of the two length scales ${hE}/{\gamma}$ and ${h}/{\Omega^{1/3}}$.

Lastly, we turn to comparing the model values for $m$ directly to the experiments. From Eq.~\ref{eqn:m} we see that $m$ depends on both the elastic properties, which affects mechanical coupling, and on the volume of the droplets, which affects the absorption. In the experiments, $h$ is controlled, but $\Omega^{1/3}$ varies from approx $\SI{300}{\micro\meter}$ to $\SI{500}{\micro\meter}$. We fit Eq.~\ref{eqn:m} directly to the data in Fig.~\ref{fig:power}, with  $E(\text{V51}) = \SI{790}{\kilo\pascal}$, $E(\text{V35}) = \SI{1230}{\kilo\pascal}$, and $E(\text{gel})=\SI{500}{\kilo\pascal}$. The values for $\Omega^{1/3}$ are given in Table~\ref{tab:m}. The model, though partially based on an empirical approach, captures the data well as shown in Table~\ref{tab:m}, for best fit parameters $E_0 = 63 \pm 8$~kPa, $\alpha=0.45 \pm 0.25$, and $\lambda(\text{gel})=0$, $\lambda(\text{V35})=-8.5 \pm 3.1$, and $\lambda(\text{V51})=-7.1 \pm 2.3$, for the gel,  high modulus, and low modulus substrates. We see that on average the calculated values of the exponent, $m_\text{th}$, agree with experiments, $m_\text{exp}$, to within about 8\%. In order to show the fit best fit lines on Fig.~\ref{fig:power}, we plot the curves for the different values of $\Omega^{1/3}$ from $\SI{300}{\micro\meter}$ to $\SI{500}{\micro\meter}$ in increments of $\SI{50}{\micro\meter}$. We emphasize that the model is empirical in nature and is intended only to describe the late-stage spreading dynamics. While it captures the dominant effects of absorption and mechanical coupling, and reflects the essential underlying physics, it remains far from a comprehensive theoretical description.

% table
\begin{table}[h!]
\centering
\caption{Experimental and theoretical values of $m$ for various film thicknesses $h$ and $\Omega^{1/3}$. Relative differences are shown as percentages.}
\label{tab:m}

% Optional: tighten or loosen horizontal padding (default ~6pt)
\setlength{\tabcolsep}{4pt}

\begin{tabular*}{\columnwidth}{@{\extracolsep{\fill}} %
  l
  S[table-format=4.0]   % h
  S[table-format=3.0]   % Ω^{1/3}
  S[table-format=1.2]   % m_exp
  S[table-format=1.2]   % m_th
  S[table-format=2.0]   % % diff
@{}}
\toprule
{Sample} &
\multicolumn{1}{c}{$h$} &
\multicolumn{1}{c}{$\Omega^{1/3}$} &
\multicolumn{1}{c}{$m_{\mathrm{exp}}$} &
\multicolumn{1}{c}{$m_{\mathrm{th}}$} &
\multicolumn{1}{c}{diff} \\
% Units row (no horizontal rule between header rows)
& {(nm)} & {($\mu$m)} & {} & {} & {\%} \\
\midrule
V35             & 14   & 450 & 0.11 & 0.12 &  8 \\
gel             & 23   & 330 & 0.13 & 0.13 &  2 \\
                & 47   & 372 & 0.13 & 0.15 &  13 \\
                & 55   & 376 & 0.15 & 0.16 &  2 \\
                & 72   & 348 & 0.15 & 0.17 &  10 \\
                & 75   & 371 & 0.15 & 0.17 &  11 \\
                & 129  & 405 & 0.17 & 0.19 &  10 \\
                & 134  & 403 & 0.16 & 0.20 &  22 \\
                & 312  & 491 & 0.18 & 0.22 &  20 \\
                & 547  & 453 & 0.22 & 0.23 &  5 \\
                & 869  & 404 & 0.23 & 0.23 &  3 \\
                & 1151 & 494 & 0.21 & 0.23 &  8 \\
                & 1427 & 487 & 0.19 & 0.23 &  15 \\
                & 1732 & 431 & 0.22 & 0.23 &  4 \\
                & 2078 & 493 & 0.20 & 0.23 &  14 \\
\midrule
V35             & 42   & 397 & 0.13 & 0.13 &  3 \\
elast.          & 115  & 379 & 0.15 & 0.15 &  4 \\
                & 220  & 398 & 0.18 & 0.15 &  18 \\
                & 232  & 371 & 0.16 & 0.15 &  11 \\
                & 369  & 449 & 0.16 & 0.14 &  8 \\
                & 619  & 453 & 0.14 & 0.14 &  2 \\
                & 987  & 340 & 0.13 & 0.13 &  0 \\
                & 1163 & 360 & 0.12 & 0.12 &  1 \\
                & 1208 & 368 & 0.12 & 0.12 &  7 \\
                & 1575 & 434 & 0.12 & 0.12 &  3 \\
\midrule
V51             & 22   & 323 & 0.13 & 0.12 &  3 \\
elast.          & 208  & 337 & 0.16 & 0.17 &  6 \\
                & 739  & 340 & 0.15 & 0.16 & 12 \\
                & 1137 & 402 & 0.14 & 0.16 & 13 \\
                & 2560 & 315 & 0.13 & 0.12 & 9 \\
\bottomrule
\end{tabular*}
\end{table}

\section*{Conclusions}

On rigid substrates, completely wetting droplets follow Tanner’s law, with the radius growing as $t^{1/10}$. Here, we show that silicone oil droplets spreading on crosslinked PDMS elastomer and gel films supported by a rigid silicon substrate deviate from this classical scaling because the droplets deform the surface and, in the case of elastomers, simultaneously swell the substrate. In contrast to prior studies, which typically examined as-prepared (i.e., swollen) semi-infinite substrates, we fabricated true elastomer thin films. By using thin films and varying the thickness $h$, we simultaneously tune the effective stiffness of the sample and the absorption potential of the elastomer substrate. Systematically varying $h$ exposes clear departures from Tanner’s law that highlight the roles of absorption into the substrate, and droplet-induced mechanical deformation of the substrate. We characterize the dynamics through the prefactor $\mathcal{C}$ and exponent $m$ assuming the typical long-time exponential growth of the radius of the droplet: $R/\Omega^{1/3} = \mathcal{C} \left( t/\tau \right)^m $. We introduce a simple, empirically motivated scaling model that captures the dependence of $\mathcal{C}$ and $m$ on film thickness, film modulus, and sample type (gel versus elastomer). The analysis identifies two mechanisms: a \textit{mechanical coupling}, which operates for both gels and elastomers, and an \textit{absorption} contribution, which is specific to elastomer films. Together, these ingredients provide a unified description of spreading on soft, swellable substrates, and offer a practical route to control wetting via substrate thickness, modulus, and permeability.

\begin{acknowledgments}
LD, BB, JRD and KDV acknowledge financial support from the Natural Science and Engineering Research Council of Canada. KDV gratefully acknowledges the Joliot chair from the ESPCI. 
\end{acknowledgments}

\section*{Author Contributions}
LD and KDV designed experiments. LD performed all experiments and analysis. ER and KDV developed the theoretical model. BB and JRD provided modulus measurements. LD wrote initial draft of manuscript and all authors contributed to writing the manuscript.

\section*{Data Availability Statement}
The datasets generated during the current study are provided herein, any further data is available from the corresponding author upon reasonable request.

\bibliographystyle{apsrev}   % or another style like unsrt, apsrev4-2, etc.
\bibliography{references}   % omit the `.bib` extension

\end{document}